\documentclass[twocolumn,aps,prd,superscriptaddress]{revtex4}

\usepackage{graphicx}
\usepackage{dcolumn}
\usepackage{amsmath}
\usepackage{epsfig}

\input babarsym.tex

\newcommand{\lumi}    {468\invfb}
\newcommand{\lumion}  {426\invfb}
\newcommand{\lumioff} {42\invfb}

\newcommand{\llll}     {\ensuremath{\ell^{-}_{1}\ell^{+}_{2}\ell^{-}_{3}}}
\newcommand{\eee}     {\ensuremath{e^-\!e^+\!e^-}}
\newcommand{\eemw}    {\ensuremath{\mu^+\!e^-\!e^-}}
\newcommand{\eemr}    {\ensuremath{\mu^-\!e^+\!e^-}}
\newcommand{\emmw}    {\ensuremath{e^+\!\mu^-\!\mu^-}}
\newcommand{\emmr}    {\ensuremath{e^-\!\mu^+\!\mu^-}}
\newcommand{\mmm}     {\ensuremath{\mu^-\!\mu^+\!\mu^-}}

\newcommand{\taulll}  {\ensuremath{\tau^{-}\!\to\llll}}

\newcommand{\dMdE}    {\ensuremath{(\Delta M_{ec}, \Delta E)}}

\def\kk2f       {\mbox{\tt KK2f}\xspace}
\def\tauola     {\mbox{\tt Tauola}\xspace}
\def\photos     {\mbox{\tt Photos}\xspace}
\newcommand{\Nobs}      {\ensuremath{N_{\rm obs}}}
\newcommand{\Nbgd}      {\ensuremath{N_{\rm bgd}}}
\newcommand{\Nul}       {\ensuremath{N_{\rm UL}^{90}}}

\newcommand{\BABARPubYear}    {09}
\newcommand{\BABARPubNumber}  {033}
\newcommand{\SLACPubNumber}   {14009}

\def\figurebox#1#2#3{%
    \def\arg{#3}%
    \ifx\arg\empty
    {\hfill\vbox{\hsize#2\hrule\hbox to #2{\vrule\hfill\vbox to #1{\hsize#2\vfill}\vrule}\hrule}\hfill}%
    \else
    {\hfill\epsfbox{#3}\hfill}%
    \fi}

\begin{document}
\begin{flushleft}
\babar-PUB-\BABARPubYear/\BABARPubNumber \\ 
SLAC-PUB-\SLACPubNumber
\end{flushleft}
\title{
{\large \bf \boldmath Limits on  $\tau$ Lepton-Flavor Violating Decays in Three Charged Leptons}}

%
\author{J.~P.~Lees}
\author{V.~Poireau}
\author{E.~Prencipe}
\author{V.~Tisserand}
\affiliation{Laboratoire d'Annecy-le-Vieux de Physique des Particules (LAPP), Universit\'e de Savoie, CNRS/IN2P3,  F-74941 Annecy-Le-Vieux, France}
\author{J.~Garra~Tico}
\author{E.~Grauges}
\affiliation{Universitat de Barcelona, Facultat de Fisica, Departament ECM, E-08028 Barcelona, Spain }
\author{M.~Martinelli$^{ab}$}
\author{A.~Palano$^{ab}$ }
\author{M.~Pappagallo$^{ab}$ }
\affiliation{INFN Sezione di Bari$^{a}$; Dipartimento di Fisica, Universit\`a di Bari$^{b}$, I-70126 Bari, Italy }
\author{G.~Eigen}
\author{B.~Stugu}
\author{L.~Sun}
\affiliation{University of Bergen, Institute of Physics, N-5007 Bergen, Norway }
% ...
\collaboration{The \babar\ Collaboration}
\noaffiliation

\date{\today}

\begin{abstract}
A search for the neutrinoless, lepton-flavor violating decay of the 
$\tau$ lepton into three charged leptons has been performed  using an 
integrated luminosity of \lumi 
collected with the \babar\ detector at the \pep2\ collider. 
In all six decay modes considered, the numbers of events found 
in data are compatible with the background expectations.
Upper limits on the branching fractions are set in the range 
$(1.8-3.3) \times10^{-8}$ at 90\% confidence level.
\end{abstract}

\maketitle


Lepton-flavor violation (LFV) involving charged leptons has 
never been observed, and stringent experimental limits 
exist \cite{brooks99, sindrum88, inami}. 
The experimental observation of neutrino oscillations~\cite{neut}
implies that, within the standard model (SM), there are amplitudes contributing
to LFV in the charged sector, although their effects must be well below the 
current experimental sensitivity \cite{pham98}.
Many descriptions of physics beyond the SM predict enhanced LFV in $\tau$ 
decays over $\mu$ decays with branching fractions  within present experimental 
sensitivities~\cite{paradisi05, babu02, brignole03}.
An observation of LFV in $\tau$ decays would be a 
clear signature of new physics, while improved 
limits will further constrain models. 


This paper reports the latest results from \babar\ 
on the search for LFV in the neutrinoless
decay \taulll{}, where $\ell_{i} = e, \mu$ \cite{cc}.
All six lepton combinations consistent with charge
conservation are considered. 
The analysis is based on data recorded 
by the \babar~ \cite{detector} detector at the \pep2\ asymmetric-energy \epem\ 
B factory operated at the SLAC National Accelerator Laboratory.
The data sample is provided by an integrated luminosity of \lumion\ recorded at 
a center-of-mass (c.m.) energy
$\sqrt{s} = 10.58 \gev$, and of \lumioff\ recorded at about
$\sqrt{s} = 10.54 \gev$.
With these conditions, the expected cross section for 
$\tau$-pair production is $\sigma_{\tau\tau} = 0.919\pm0.003$ nb 
\cite{tautau}, corresponding to a data sample of about 430 million $\tau$-pairs.


Charged-particle (track) momenta are measured with a 5-layer
double-sided silicon vertex tracker and a 40-layer helium-isobutane 
drift chamber inside a 1.5 T superconducting solenoid magnet.
An electromagnetic calorimeter consisting of 6580 CsI(Tl) 
crystals is used to measure electron and photon energies,
a ring-imaging Cherenkov detector is used to identify
charged hadrons, 
and the instrumented magnetic flux return (IFR) is used to identify muons.
About half of the data sample under study was recorded with the IFR 
instrumented with resistive plate chambers (RPC). During the second half of the
data taking period  most RPCs
were replaced by limited streamer tubes in the barrel section of the IFR. 

A Monte Carlo (MC) simulation of lepton-flavor violating $\tau$ decays
is used to estimate the signal efficiency and optimize the search.
Simulated $\tau$-pair events including higher-order radiative
corrections are generated using \kk2f \cite{kk}
with one $\tau$ decaying to three leptons with a uniform three-body 
phase space distribution, while the other $\tau$ decays 
according to measured rates \cite{PDG} simulated with \tauola \cite{tauola}.
Final-state radiative effects are simulated for all decays 
using \photos \cite{photos}.
The detector response is simulated with \mbox{\tt GEANT4}~\cite{geant}.

The signature for \taulll{} is a set of three charged 
particles, each identified as either an $e$ or a $\mu$,
with an invariant mass and energy equal to that of the parent $\tau$ lepton.
Events are preselected requiring four reconstructed tracks 
and zero net charge,
selecting only  tracks pointing toward a common region consistent with 
\tautau production and decay. 
The polar angles of all four tracks in the laboratory
frame are required to be within the calorimeter acceptance range, to ensure
 good particle identification. 
The event is divided into two hemispheres 
in the \epem\ c.m. frame 
using the plane containing the interaction point and perpendicular to the 
thrust axis, as calculated from the observed tracks and neutral energy 
deposits.
The signal hemisphere must contain exactly three tracks (3-prong) with an
invariant mass less than 3.5\gevcc, 
while the other hemisphere must contain exactly one (1-prong) track, and may
contain also neutral energy deposits. 
In order to reduce backgrounds coming from photon conversions 
we  require that the two couples of oppositely charged tracks in the 
3-prong hemisphere have an invariant mass, 
calculated using  electron mass hypothesis 
for the tracks, larger than than 20\mevcc, or 30\mevcc for \eee and 
\emmr.


With respect to our previous result~\cite{kolb}, this analysis relies on 
significantly improved particle identification (PID) techniques for both 
$\mu^{\pm}$ and $e^{\pm}$.
Electrons are identified applying a multivariate algorithm which uses as input 
the ratio of calorimeter energy to track momentum $(E/p)$, the ionization 
energy loss in the tracking system $(\dedx)$, and the shape of the shower
in the calorimeter.
Muon identification exploits a bagged decision tree (BDT)\cite{BDT} 
algorithm, which uses as input the number of hits in the IFR,
the number of interaction lengths traversed,
and the energy deposition in the calorimeter.
Since $\mu^{\pm}$ with momenta less than $500\mevc$ do not penetrate enough 
into the IFR to provide useful information, the BDT also uses information 
obtained from the inner trackers
to maintain
a very low $\pi-\mu$ misidentification probability with high selection 
efficiencies.
The electron and muon identification efficiencies
are measured to be 91\% and 77\% respectively. 
The probability for a $\pi^{\pm}$ to be misidentified as an $e^{\pm}$ in 
3-prong $\tau$ decays is 2.4\%, while the probability to be misidentified
as a $\mu^{\pm}$ is 2.1\%. 

The quantity
$\Delta E \equiv E^{\star}_{\mathrm{rec}} - E^{\star}_{\mathrm{beam}}$
is defined, where $E^{\star}_{\mathrm{rec}}$ is the total energy of the system 
observed in the 3-prong hemisphere and $E^{\star}_{\mathrm{beam}}$
is the beam energy (the superscript $\star$ indicates quantities measured in 
the c.m. frame).
We define
$\Delta M_{ec} \equiv M_{\mathrm{ec}} - m_{\tau}$ with 
$M_{\mathrm{ec}}^{2}\equiv E^{\star \, 2}_{beam}/c^{4} - |\vec{p}_{3l}^{\, \star}|^{2}/c^{2}$, 
where $|\vec{p}_{3l}^{\, \star}|^{2}$ 
is the squared momentum of the 3-prong system,
$m_{\tau}=1.777\gevcc$ is the $\tau$ mass \cite{PDG}, and the energy 
constrained momentum of the 3-prong system, 
$|\vec{p}_{3l}^{\star}|$, 
is obtained from a kinematic
fit: the fit requires the $\tau$ energy measured in the c.m. to be 
$\sqrt{s/2}$, 
taking into account the errors on the reconstructed track parameters 
and the beam energy measurement.

\begin{figure}
\includegraphics[scale=0.50]{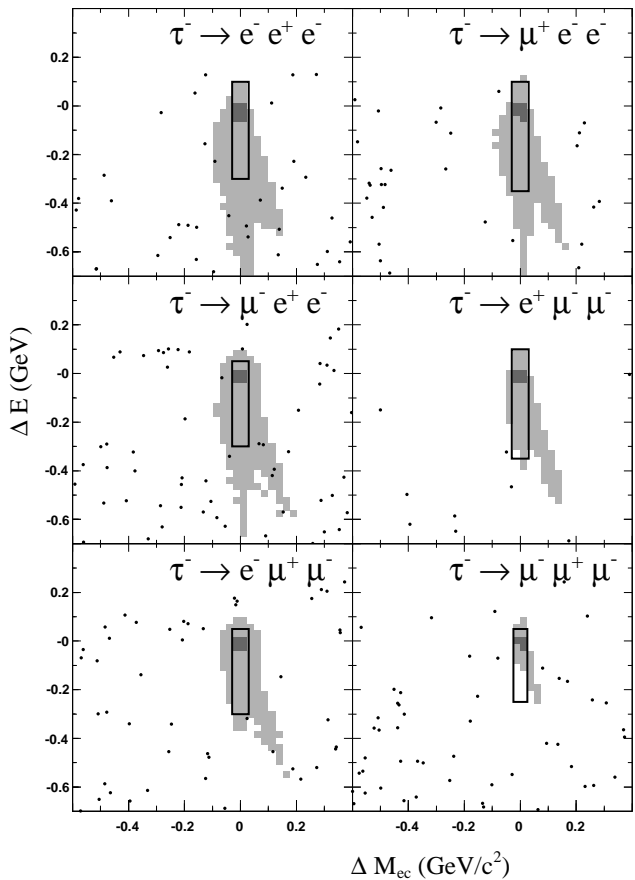}
\caption{Data events (dots) in the large box of the \dMdE\, as defined in 
the text, for the six $\tau$ decay channels after all
selection is applied. 
The solid black lines are the boundaries, for each channel, of the signal
region. The dark and light shadings 
represent the 50\% and 90\% signal contours respectively.}
\label{fig1}
\end{figure}

The signal distributions in the \dMdE\ plane (see Fig.~\ref{fig1})
are broadened by detector resolution and radiative effects.
In all decay modes, 
the radiation of photons from the incoming \epem\ particles
and from the outgoing $\tau$ decay products 
leads to a tail at low values of $\Delta E$.
Radiation from the final-state leptons, 
which is more likely for electrons than for muons, 
produces a tail at high values of $\Delta M_{ec}$ as well.
Signal regions (SR) in the \dMdE\ plane are optimized in order to obtain the
smallest expected upper limit (UL) when no LFV signal is present.
The expected ULs are estimated using MC
simulations and data control samples, instead of candidate 
signal events.
The upper right corner of
the signal region in the \dMdE\ plane, in units of $(\mevcc, \mev)$,
is fixed at $(30, 50)$ for \eemr\ and \emmr\, and at $(30, 100)$ for the 
other four channels.
The lower left corner is at 
$(-30, -300)$ for the \eee, \eemr, and \emmr decay modes, 
$(-30, -350)$ for \eemw\ and \emmw, 
and $(-25, -200)$ for \mmm. 
Fig.~\ref{fig1} shows the observed data in the \dMdE\ plane, 
along with the signal region boundaries and the expected signal 
distributions.
To avoid biases, a blind analysis procedure was followed,
with the number of events in the SR remaining unknown
until the selection criteria were finalized and all cross-checks were 
performed.

Each track present in the signal hemisphere must be identified as either 
a muon or an electron, depending on the channel under study. For the channels
where two tracks of the same charge sign can be either an electron or a muon 
(i.e. \eemr\ and \emmr), it is possible that both tracks satisfy both electron 
and muon PID selectors: in these rare cases we measure $\Delta M_{ec}$ and 
$\Delta E$ in both mass hypotheses. For all events showing this behavior only
one of the two combinations falls in the 
large box (LB) of the \dMdE\ plane, defined as the region  
lying between $-600$ and $400$\mevcc in $\Delta M_{ec}$ and
$-700$ and $400$\mev in $\Delta E$.

The PID requirements strongly suppress background, but further selection is 
applied:
for all decay modes, the momentum of the 1-prong track is required 
to be less than 4.8 \gevc in the c.m.\ frame.
The 1-prong side $\tau$ mass is approximately reconstructed from
the 4-momentum obtained by adding the 1-prong track, the neutral 
energy deposits in the 1-prong hemisphere, and the missing 3-momentum of the 
event, assuming a zero mass as is appropriate if just a single neutrino 
is missing. This invariant mass
is required to be in the range 
$0.2-3.0$ \gevcc 
for \eee{}, \eemr{}, and \emmr{}, and in the range of~$0.1-3.5$ \gevcc{}
for \eemw{}, \emmw{}, and \mmm.
To suppress Bhabha backgrounds we reject events where any oppositely 
charged track 
pair has an invariant mass compatible with a photon conversion when assigning 
the electron mass to the two tracks. $M_{e^{+}e^{-}}$ is required to be $
<200$\mevcc{} 
for  all channels except for \emmw\ where $M_{e^{+}e^{-}}<300$\mevcc{} is 
required.
For the \eee\ and \emmr\ decay modes, the charged particle
in the 1-prong hemisphere is required to be matched to an energy deposit in the
calorimeter inconsistent with an electron, and must not be identified as an
electron, while for the \eemw, \eemr, and \mmm\ decay modes
this track must not be identified as a muon.
For the \eee\ and \emmr\ decay modes,
the missing momentum of the event should be greater than
$300$\mevc{}, for \emmr\ and \mmm\ this should be more than 
$200$\mevc{} and for \eemw\ and \emmw this lower limit is 
set at $100$\mevc.
For the \eee\ and \eemw\ channels the cosine of the angle between
the direction of the sum of the three signal track momenta and
the direction of the 1-prong track momentum ($\theta_{13}$), is required
to satisfy  $\cos(\theta_{13})>-0.995$ and $\cos(\theta_{13})>-0.997$ 
respectively, to further reduce Bhabha contributions.

The backgrounds still contaminating the sample have been identified in three
broad categories: low multiplicity $q\bar{q}$ events
(comprising both continuum light quark pairs and $c\bar{c}$ pairs),
 QED events (Bhabha or $\mu^{+}\mu^{-}$ depending on the particular channel),
 and SM $\tau^{+}\tau^{-}$ events.
These three background classes have distinctive distributions
in the \dMdE\ plane.
The  $q\bar{q}$ events tend to populate the plane uniformly,
while QED backgrounds fall in a narrow band
at positive values of $\Delta E$, and $\tau^{+}\tau^{-}$  backgrounds
are restricted to negative values of both $\Delta E$ and $\Delta M_{ec}$ due to
the presence of at least one undetected neutrino.
The possible  background contribution arising from 
two-photon processes has been studied on a data control sample,
as discussed in the following, and it is found to be
negligible.

The expected background rates for each decay mode are determined by
fitting a set of probability density functions (PDFs) to the
observed data in the grand sideband (GS) region of the \dMdE\ plane as was done
in the previous published analysis \cite{kolb}.
The GS region covers the same region as the LB but does not include the SR.
The functional forms of the PDFs are the same as in \cite{kolb}.
For the \qqbar background, a two-dimensional PDF is
constructed from the product of two PDFs, $P_{M^\prime}$ and %
$P_{E^\prime}$, where %
$P_{M^\prime}(\Delta M^\prime)$ is a bifurcated Gaussian and %
$P_{E^\prime}(\Delta E^\prime) = 
(1-x/\sqrt{1+x^2})(1+a x+b x^2+c x^3)$ with $x=(\Delta E^\prime-d)/e$.
The $(\Delta M^\prime, \Delta E^\prime)$ axes have
been slightly rotated from $(\Delta M_{ec}, \Delta E)$ to take into
account the observed correlation between $\Delta E$ and $\Delta M_{ec}$
for the distribution.
For the \tautau\ background PDF, the function $P_{M^{\prime\prime}}(\Delta M^{\prime\prime})$
is the sum of two Gaussians with common mean, while the functional form 
of $P_{E^{\prime\prime}}(\Delta E^{\prime\prime})$ is the same as that for the \qqbar\ PDF.
To properly model the wedge-shaped distribution due to the kinematic
limit in tau decays, a coordinate transformation of the form
$\Delta M^{\prime\prime}=\textrm{cos}\beta_1\Delta M_{ec}+
\textrm{sin}\beta_1\Delta E$ and 
$\Delta E^{\prime\prime}=\textrm{cos}\beta_2\Delta E-
\textrm{sin}\beta_2\Delta M_{ec}$ is performed. 

QED backgrounds represent one of the major sources of backgrounds for
\eee{}, \eemw{}, \eemr{}, and \emmr. To study this background category,  
specially selected control
samples obtained from data were produced. 
Two different methods were used to extract QED control samples: for \eee, 
\eemw, and \emmr\ channels the sample was produced selecting events passing all
selection requirements, except the lepton veto in the tag side, and requiring 
the track in the tag-side to be identified as a muon (\eemw\ channel) or as
an electron (for the other two channels). To obtain a large enough sample for 
the 
\eemr\ channel we selected events where a muon is present in the tag-side, and
the reconstructed mass in the tag-side is between 0.5\gevcc\  and 2.5\gevcc, 
and
the momentum of the tag-side particle is required to be larger than 4.8\gevc.
To fit these control samples,
an analytic PDF is constructed from
the product of a Crystal Ball function \cite{CBF} in $\Delta E'$ 
and a third-order polynomial in $\Delta M'$, where again the
$(\Delta M', \Delta E')$ axes have been rotated slightly
from \dMdE\ to fit the observed distribution.

The expected background rate in the SR is obtained
by an unbinned maximum likelihood fit to the data in the GS region, with 
the shapes of the three background PDFs fixed by making an unbinned 
likelihood fit to the MC and the control samples.
The PDF shape determinations and background fits are performed
separately for each of the six decay modes.
Cross-checks of the background estimation are performed by 
considering the numbers of events expected and observed in 
sideband regions immediately neighboring the signal region for 
each decay mode.

The efficiency of the selection for signal events is estimated with the 
MC simulation of signal LFV events.
The efficiency of signal MC passing preselection requirements varies between
45\% and 49\%.
The total efficiency for signal events to be found
in the signal region is shown in Table~\ref{tab:results} for
each decay mode and ranges from 6.4\% to 12.7\%.
This efficiency includes the 85\% branching fraction for 1-prong 
$\tau$ decays. 
With respect to the previous analysis, improvements in particle ID,
in tracking algorithms and in selection criteria allowed us to obtain 
higher signal efficiencies along with a reduction of the expected backgrounds 
thus improving the UL sensitivity.

\begin{table}
\begin{center}
\caption{Efficiencies, numbers of expected background events (\Nbgd),
expected branching fraction upper limits at 90\% CL (UL$_{90}^{\rm exp}$),
numbers of observed events (\Nobs), and observed branching fraction upper 
limits 
at 90\% CL (UL$_{90}^{\rm obs}$) for each decay mode. All upper limits are in
units of $10^{-8}$.
}

\begin{tabular}{lcccccc}
\hline\hline
Mode & Eff. [\%] && \Nbgd  & UL$_{90}^{\rm exp}$ & \Nobs & \rule{0pt}{13pt}UL$_{90}^{\rm obs}$ \\
\hline
\eee  &$ 8.6  \pm 0.2 $&&$ 0.12 \pm 0.02 $&$ 3.4  $&$ 0 $&$ 2.9 $\\ 
\eemr &$ 8.8  \pm 0.5 $&&$ 0.64 \pm 0.19 $&$ 3.7  $&$ 0 $&$ 2.2 $\\
\eemw &$ 12.7 \pm 0.7 $&&$ 0.34 \pm 0.12 $&$ 2.2  $&$ 0 $&$ 1.8 $\\
\emmw &$ 10.2 \pm 0.6 $&&$ 0.03 \pm 0.02 $&$ 2.8  $&$ 0 $&$ 2.6 $\\
\emmr &$ 6.4  \pm 0.4 $&&$ 0.54 \pm 0.14 $&$ 4.6  $&$ 0 $&$ 3.2 $\\
\mmm  &$ 6.6  \pm 0.6 $&&$ 0.44 \pm 0.17 $&$ 4.0  $&$ 0 $&$ 3.3 $\\
\hline
\hline
\end{tabular}
\label{tab:results}
\end{center}
\end{table}

Uncertainties in signal efficiency 
estimation and in the number of the expected events in the SR obtained by 
the fit affect the final result.
The systematic uncertainties from PID dominate the error on the efficiency.
They are estimated on data control samples, looking at the 
discrepancies between data and MC samples, by measuring the average spread for
tracks with the same kinematic properties. These uncertainties vary between 
a relative error of 1.8\% for \eee\ and 7.8\% for \mmm. 
The modeling of the tracking efficiency contributes an additional 1\% relative
uncertainty.
All other sources of uncertainty in the signal efficiency are found to be 
smaller than $1.0 \%$, including the statistical limitation of the MC 
signal samples, the modeling of higher-order radiative effects, 
track momentum resolution, trigger performance, observables used in 
the selection criteria, and knowledge of the tau 1-prong branching 
fractions.

The systematic uncertainty due to errors in background estimation is
determined from fits to data in the GS region. In addition to varying 
PDF parameters by their uncertainties, alternative functional forms are used
to determine the uncertainty on the expected background yield in the SR.
The total errors on the background estimates 
are reported in Table~\ref{tab:results}.
Systematics coming from unsimulated background contributions, such as 
two-photon processes, are checked using background enriched control samples. 
Two-photon processes are characterized by a small transverse momentum, so the
control samples were produced selecting events with a transverse momentum 
smaller than 0.2\gevc, and with the momentum of the tag-side track in the 
center of mass smaller than 4.0\gevc. The uncertainties introduced by 
two-photon processes and unsimulated backgrounds are found to be negligible. 

Background expectations (\Nbgd) and the number of observed events (\Nobs)
are shown in Table~\ref{tab:results}.
No events are observed in the SR for any of the modes and we place 90\% 
confidence level (CL)
ULs on the branching fractions using
UL$_{90}^{\rm exp} = \Nul/(2 \varepsilon \L \sigma_{\tau\tau})$, where $\Nul$
is the 90\% CL UL for the number 
of signal events when \Nobs\ events are observed with \Nbgd\ background 
events expected.
The values $\varepsilon$, $\L$, and $\sigma_{\tau\tau}$ are the
selection efficiency, luminosity, and \tautau cross section, respectively.
The uncertainty on the product $\L \cdot \sigma_{\tau\tau}$ is 0.9\%. 
The branching fraction ULs are calculated, with 
all uncertainties included, using the technique of 
Cousins and Highland \cite{cousins92} following the implementation of 
Barlow \cite{barlow02}. 
The  expected average upper limit UL$_{90}^{\rm exp}$, 
defined as the mean UL expected in the background-only hypothesis, 
is included in Table~\ref{tab:results}.
The 90\% CL ULs on the \taulll\ branching fractions 
are in the range $(1.8-3.3)\times10^{-8}$.
These limits supersede the previous \babar\ analysis \cite{kolb}, and are 
compatible with the latest Belle limits \cite{bellelll}.

We are grateful for the excellent luminosity and machine conditions
provided by our \pep2\ colleagues, 
and for the substantial dedicated effort from
the computing organizations that support \babar.
The collaborating institutions wish to thank 
SLAC for its support and kind hospitality. 
This work is supported by
DOE
and NSF (USA),
NSERC (Canada),
CEA and
CNRS-IN2P3
(France),
BMBF and DFG
(Germany),
INFN (Italy),
FOM (The Netherlands),
NFR (Norway),
MES (Russia),
MICIIN (Spain),
STFC (United Kingdom). 
Individuals have received support from the
Marie Curie EIF (European Union),
the A.~P.~Sloan Foundation (USA)
and the Binational Science Foundation (USA-Israel)


\begin{thebibliography}{0}
\bibitem{cc}
Throughout this paper, charge conjugate decay modes are implied.
\end{thebibliography}
\end{document}